\documentclass[usenatbib,useAMS,fleqn,a4paper]{mnras}
\usepackage{amssymb}
\usepackage[T1]{fontenc}
\usepackage{ae,aecompl}
\usepackage{longtable}

\usepackage{graphicx}

\date{Accepted XXX. Received YYY; in original form ZZZ}
\pubyear{2016}

\begin{document}
\label{firstpage}
\pagerange{\pageref{firstpage}--\pageref{lastpage}}

\title[An exoplanet in the Solar System?]
  {Is there an exoplanet in the Solar System?}

\author[Mustill, Raymond, \& Davies]
  {Alexander J. Mustill$^1$\thanks{E-mail: alex@astro.lu.se}, 
    Sean N. Raymond$^{2,3}$
    and Melvyn B. Davies$^1$\\
    $^1$Lund Observatory, Department of Astronomy \& Theoretical Physics, 
    Lund University, Box 43, SE-221 00 Lund, Sweden\\
    $^2$CNRS, Laboratoire d'Astrophysique de Bordeaux, UMR 5804, 
    F-33270, Floirac, France\\
    $^3$Univ. Bordeaux, Laboratoire d'Astrophysique de Bordeaux, 
    UMR 5804, F-33270, Floirac, France}
    
\maketitle

\begin{abstract}
  We investigate the prospects for the capture of 
  the proposed Planet~9 
  from other stars in the Sun's birth cluster. 
  Any capture scenario must satisfy three conditions: the encounter 
  must be more distant than $\sim150$\,au to avoid perturbing the 
  Kuiper belt; the other star must have a wide-orbit 
  planet ($a \gtrsim 100$\,au); 
  the planet must be captured onto an appropriate orbit to sculpt 
  the orbital distribution of wide-orbit Solar System bodies. Here we use N-body 
  simulations to show that these criteria may be simultaneously satisfied. 
  In a few percent of slow close encounters in a cluster, 
  bodies are captured onto heliocentric, Planet~9-like orbits. 
  During the $\sim100$\,Myr cluster phase, 
  many stars are likely to host planets on highly-eccentric 
  orbits with apastron distances beyond 100\,au if Neptune-sized planets 
  are common and susceptible to 
  planet--planet scattering. 
  While the existence of 
  Planet~9 remains unproven, we consider capture from one of the Sun's 
  young brethren a plausible route to explain such an object's 
  orbit. Capture appears to predict a large population of 
  Trans-Neptunian Objects (TNOs) whose orbits are aligned with the captured 
  planet, and we propose that 
  different formation mechanisms will be distinguishable based on 
  their imprint on the distribution of TNOs.
\end{abstract}


\section{Introduction}

Recent speculation suggests that the outer Solar System hosts a ``Planet~9'' 
of several Earth masses or greater. \cite{TrujilloSheppard14}, in 
announcing the discovery of an unusual trans-Neptunian object (TNO) of high 
perihelion (2012~VP$_{113}$), noticed a clustering of the argument of 
perihelion of bodies lying beyond $\sim150$\,au, and attributed this 
to a hypothetical super-Earth body lying at several hundred au whose 
gravity dominates over the perihelion precession induced by the known 
planets that would cause an orbital de-phasing over 100s of Myr. This 
argument has been developed by \cite{dlFMdlFM14}, who proposed two 
distant planets to explain further patterns in the distributions of 
orbital elements. \cite{Malhotra+16} point out mean-motion commensurabilities 
between the distant TNOs, which they trace back to a hypothetical body at 
$\sim665$\,au whose resonant perturbations on the TNOs would lead to 
their apsidal confinement. Earlier work was summarized and extensively 
developed by \cite{LykawkaMukai08}, who favoured a sub-Earth mass 
embryo at $100-200$\,au.

Current interest has been fomented by 
\cite{BatyginBrown16}, who show numerically and analytically how the 
apsidal and nodal clustering of the distant TNOs arises as a result of 
resonant and secular dynamical effects from a distant perturber. They 
identify a range of semi-major axes ($400-1500$\,au) and eccentricities 
($0.5-0.8$) for which a distant planet can explain the orbital elements 
of the distant TNOs, refined to a roughly triangular region in $a-e$ 
space in a follow-up study, with $a\in[300,900]$\,au and $e\in[0.1,0.8]$
 \citep[][and see Figure~\ref{fig:a-e}]{BrownBatygin16}. This range of semi-major axes is more distant than 
proposed by \cite{TrujilloSheppard14} and \cite{dlFMdlFM14}, but brackets 
the resonant perturber of \cite{Malhotra+16}. The latter authors favour a 
lower eccentricity for Planet~9, but as the range proposed by 
\cite{BrownBatygin16} is backed up by multi-Gyr numerical simulations, we 
adopt their ranges of $a$ and $e$ as orbital elements of Planet~9 in this 
paper. Unfortunately, observations currently do not constrain the 
possible orbit very strongly. Sub-Neptune mass bodies can exist undetected 
in electromagnetic emission 
at several hundred au \citep{Luhman14,LinderMordasini16,Ginzburg+16}, and 
\cite{Fienga+16} show that a dynamical analysis of Cassini ranging data may 
rule out some ranges of 
orbital phase for a Planet~9, although they restricted their analysis to 
only a single choice of $a$ and $e$. Henceforth, we consider the whole of 
the parameter space identified by \cite{BrownBatygin16} to be viable.

In this paper we investigate how the Solar System might have come to host 
a wide-orbit eccentric body such as Planet~9, a class of object we refer to 
as ``Novenitos''. A number of lines of evidence 
suggest that the Sun formed in a sizeable cluster of a few thousand stars 
\citep[see][for reviews]{Adams10,Pfalzner+15}, and previous dynamical 
studies have shown that orbiting bodies at large radii can be transferred 
between stars in the slow ($\sim1$\,km\,s$^{-1}$) close encounters typical 
in open clusters \citep{ClarkePringle93,KenyonBromley04,
MorbidelliLevison04,Pfalzner+05,Levison+10,Malmberg+11,Belbruno+12,Jilkova+15}; 
and we show that it is indeed possible for the 
Sun to have captured such a planet from another star in a close encounter 
in its birth cluster. Our study is complementary to the recent work of 
\cite{LiAdams16}, who also identify capture in a cluster as a possible source 
for Planet~9. Whereas these authors consider the capture of planets initially on 
circular or moderately-eccentric orbits, we focus on a scenario in which 
the Sun captures a highly-eccentric planet with a semi-major axis of several hundred 
au but a pericentre of $\sim10$ au. In Section~\ref{sec:flyby} we present 
our simulations for the capture of eccentric planets by the Sun; we show how 
suitable source planets may exist on highly eccentric orbits around 
their parent star for many Myr 
during eras of planet--planet scattering in Section~\ref{sec:scatter}; and we 
discuss our results in Section~\ref{sec:discuss}.

\vspace{-9mm}

\section{Capture of Novenitos in a flyby}

\label{sec:flyby}

\begin{figure}
  \includegraphics[width=.5\textwidth]{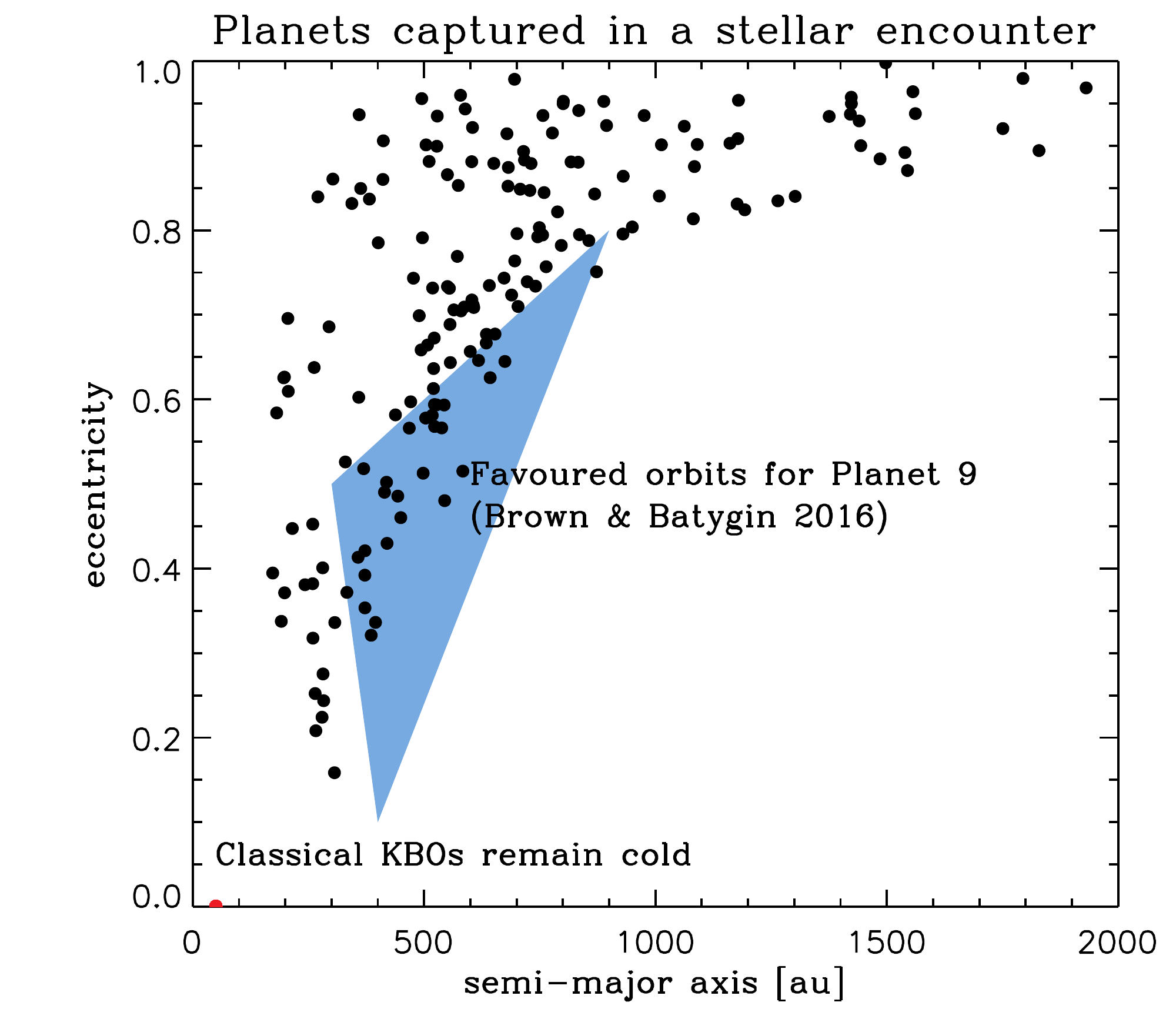}
  \caption{Semi-major axes and eccentricities of particles after a flyby 
    with $M_\star=0.1\mathrm{M}_\odot$, $b=1250$\,au, 
    $v_\mathrm{inf}=5\times10^{-4}$\,au\,d$^{-1}$, and initial orbits 
    of $a=100$\,au, $q=10$\,au around the star. The orbits of captured 
    particles around the Sun are shown in black, while the blue box 
    marks the range for Novenitos, following \protect\cite{BrownBatygin16}. In red, 
    at the lower left corner, are the post-flyby orbits of the 
    barely-perturbed Kuiper Belt Objects, initially on circular orbits 
    at 50\,au from the Sun.}
  \label{fig:a-e}
\end{figure}

We first consider the likelihood of the capture of a wide-orbit planet 
by the Sun in a close encounter with another star. Capture can occur when 
the initial orbital velocity of the planet around its original host 
and the velocity of the encounter are comparable, which leaves the 
planet with a similar orbital velocity around its new star. For the postulated orbit of Planet~9
of $\sim500$\,au, this suggests a similarly wide orbit around the original 
host (unless the host is low mass, in which case smaller orbits become favoured) 
and an encounter velocity of $\sim1$\,km\,s$^{-1}$. Given this velocity, 
we constrain the impact parameter by requiring that the cold classical Kuiper 
Belt not be disrupted during the flyby. This requires a perihelion separation 
greater than $\sim150$\,au \citep{KobayashiIda01,Breslau+14}, 
or an impact parameter greater than 500\,au, 
depending on the mass of the original host. For transfer of material between 
stars, the minimum separation must also be at most roughly three times the semi-major axis of 
the orbiting bodies \citep{Pfalzner+05}, meaning a perihelion separation $\lesssim1500$\,au 
for the close encounter. Fortunately for the capture hypothesis, these 
encounters occur remarkably frequently in clusters of a few hundred stars 
or more: \cite{Malmberg+07,Malmberg+11} found that 
only $\sim20\%$ of Solar-mass stars in a cluster of $N=700$ avoid 
a close encounter within 1000\,au, and the mean minimum perihelion distance is 
$\sim250$\,au; similarly, \cite{Adams+06} found an encounter rate of 0.01 
encounters within $\sim300$\,au per 
star per Myr in an $N=1000$ subvirial cluster.

\begin{table}
  \centering
  \caption{Parameter choices for our flyby simulations.}
  \label{tab:params}
  \begin{tabular}{lc}
    Parameter & Values\\
    \hline
    Mass of intruder $M_\star$ & $\{0.1,0.2,0.5,1.0,1.5\}\mathrm{\,M}_\odot$\\
    Impact parameter $b$       & $\{500,750,1000,1250\}$\,au\\
    Encounter velocity $v_\mathrm{inf}$ & $\{0.5,1.0\}\times10^{-3}$\,au\,d$^{-1}$\\
    Initial planet semimajor axis $a$ & $\{50,100,200,400,800\}$\,au\\
    Initial planet pericentre $q$ & $\{1,10\}$\,au
  \end{tabular}
\end{table}

\begin{figure}
  \includegraphics[width=0.5\textwidth]{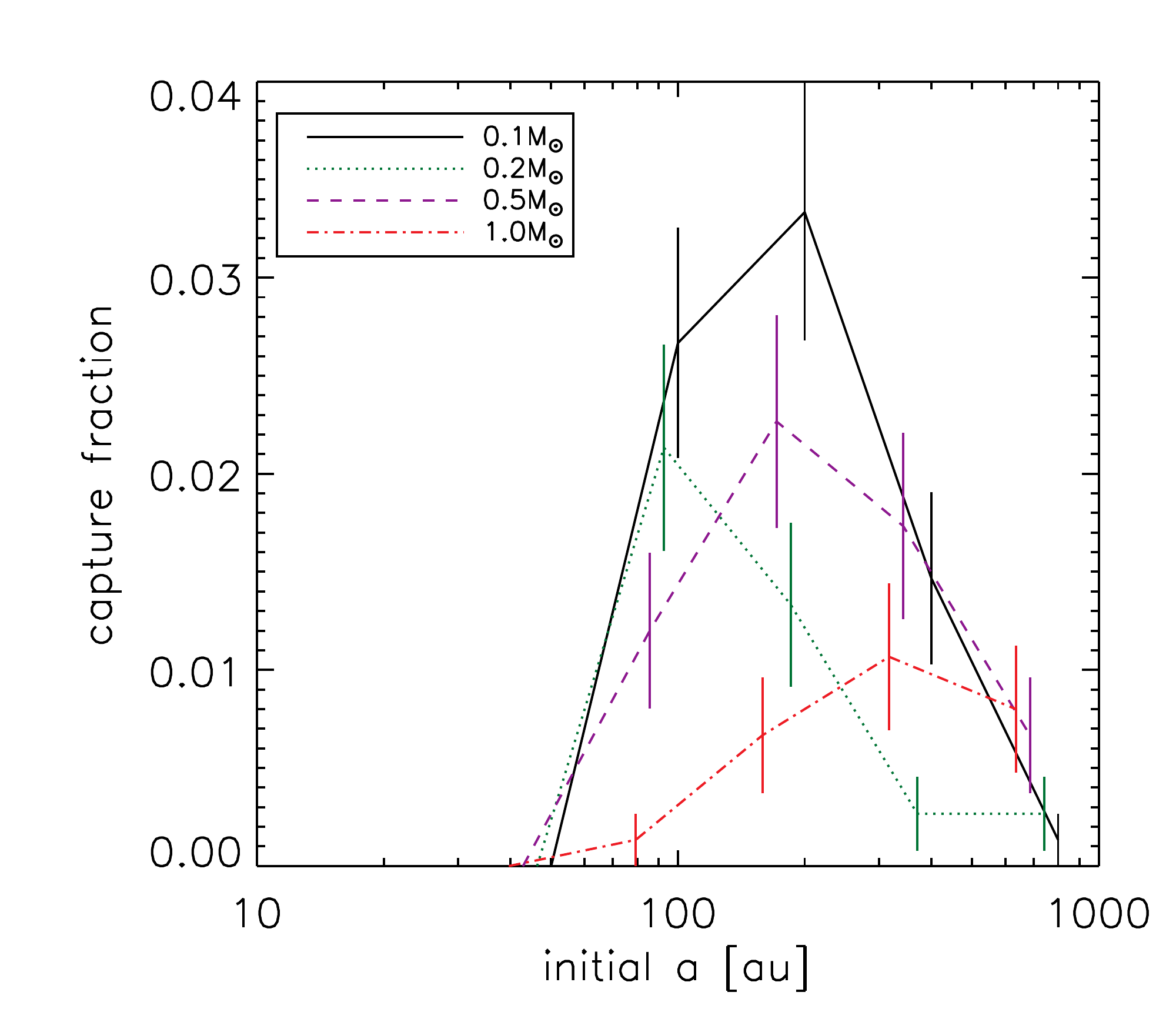}
  \caption{Fractions of particles captured into Novenito orbits for 
    objects on orbits of different initial semi-major 
    axes around stars of different masses. Other parameters are fixed at 
    bodies' initial pericentre $q=10$\,au, impact parameter $b=1000$\,au, 
    $v_\mathrm{inf}=5\times10^{-4}$\,au\,d$^{-1}$. Note that a $1.5\mathrm{\,M}_\odot$ 
    star with these encounter parameters would disrupt the Kuiper Belt. 
    Error bars show the $1\sigma$ range of the posterior distribution from inverting 
    the binomial sampling distribution. Symbols offset for clarity.}
  \label{fig:cuts}
  \end{figure}

The above considerations thus define a broad parameter space for capture 
with $v_\mathrm{inf}\sim1$\,km\,s$^{-1}\approx5.8\times10^{-4}$\,au\,d$^{-1}$, 
impact parameter $b\sim1000$\,au, 
semimajor axis $a\sim$ 500\,au. For the mass of the original host $M_\star$ we 
consider a range from $0.1$ to $1.5\mathrm{\,M}_\odot$, covering a wide 
range of known planet hosts. For the eccentricity of Planet~9's original 
orbit, we focus on very high values corresponding to pericentres of 
$1-10$\,au, consistent with the aftermath of a phase of strong planet--planet 
scattering as we describe in section~\ref{sec:scatter}. Our parameter choices 
are listed in Table~\ref{tab:params}. We explore this 
parameter space with N-body integrations using the \textsc{Mercury} package 
\citep{Chambers99}. We use the conservative BS algorithm to integrate 
the trajectories of the Sun and an intruder; the latter is surrounded by 
an isotropic swarm of massless test particles (750 per integration). The intruder 
begins at 10\,000\,au with a velocity $v_\mathrm{inf}$, and the system is 
integrated until the intruder attains a heliocentric distance of 20\,000\,au, at 
which distance bodies are removed from the integration. For each integration, we 
count the number of particles captured onto bound orbits around the Sun 
(imposing a cut-off of $a=5000$\,au to reject particles which spuriously remain 
bound after removal of the original host), as well as the number 
captured onto Novenito orbits of \cite{BrownBatygin16}.

We find that the Sun can capture bodies from the intruding star for many 
combinations of parameters. 
Examples of the final orbital elements of captured bodies are shown in 
Figure~\ref{fig:a-e}. 
Our highest capture rate is $44\%$, attained for $M_\star=0.5\mathrm{\,M}_\odot$, 
$b=750$\,au, $a=800$\,au, $v_\mathrm{inf}=5\times10^{-4}$\,au\,d$^{-1}$. However, of 
these captured particles, only 2\% (1\% of the total) 
have orbital elements suitable for a 
Novenito. Other parameter combinations give higher 
rates of capture into Novenito orbits, 
reaching almost $4\%$ for $M_\star=0.1\mathrm{\,M}_\odot$,
$b=1250$\,au, $a=100$\,au, $v_\mathrm{inf}=5\times10^{-4}$\,au\,d$^{-1}$. 
While this is our most successful simulation, capture rates of a few percent 
are attained for a much wider range of parameters. Cuts 
through the parameter space are shown in Figure~\ref{fig:cuts}. 
Our full results are given in Table~A1 of the online version of this paper.

We also ran simulations to check that the Sun's Kuiper Belt would not be disrupted 
in such an encounter. For these we distributed around the Sun 750 particles at 
$a=50$\,au, $e=0$, with isotropic inclinations, and verified that eccentricities 
remained low ($\lesssim0.1$) after the flyby. These simulations ruled out impact 
parameters of $b=1000$\,au and below for a $1.5\mathrm{\,M}_\odot$ intruder, down 
to $b=500$\,au and below for a $0.1\mathrm{\,M}_\odot$ intruder, with 
$v_\mathrm{inf}=5\times10^{-4}$\,au\,d$^{-1}$.

\vspace{-5mm}

\section{Source populations of Novenitos}

\label{sec:scatter}

\begin{figure}
  \includegraphics[width=.5\textwidth]{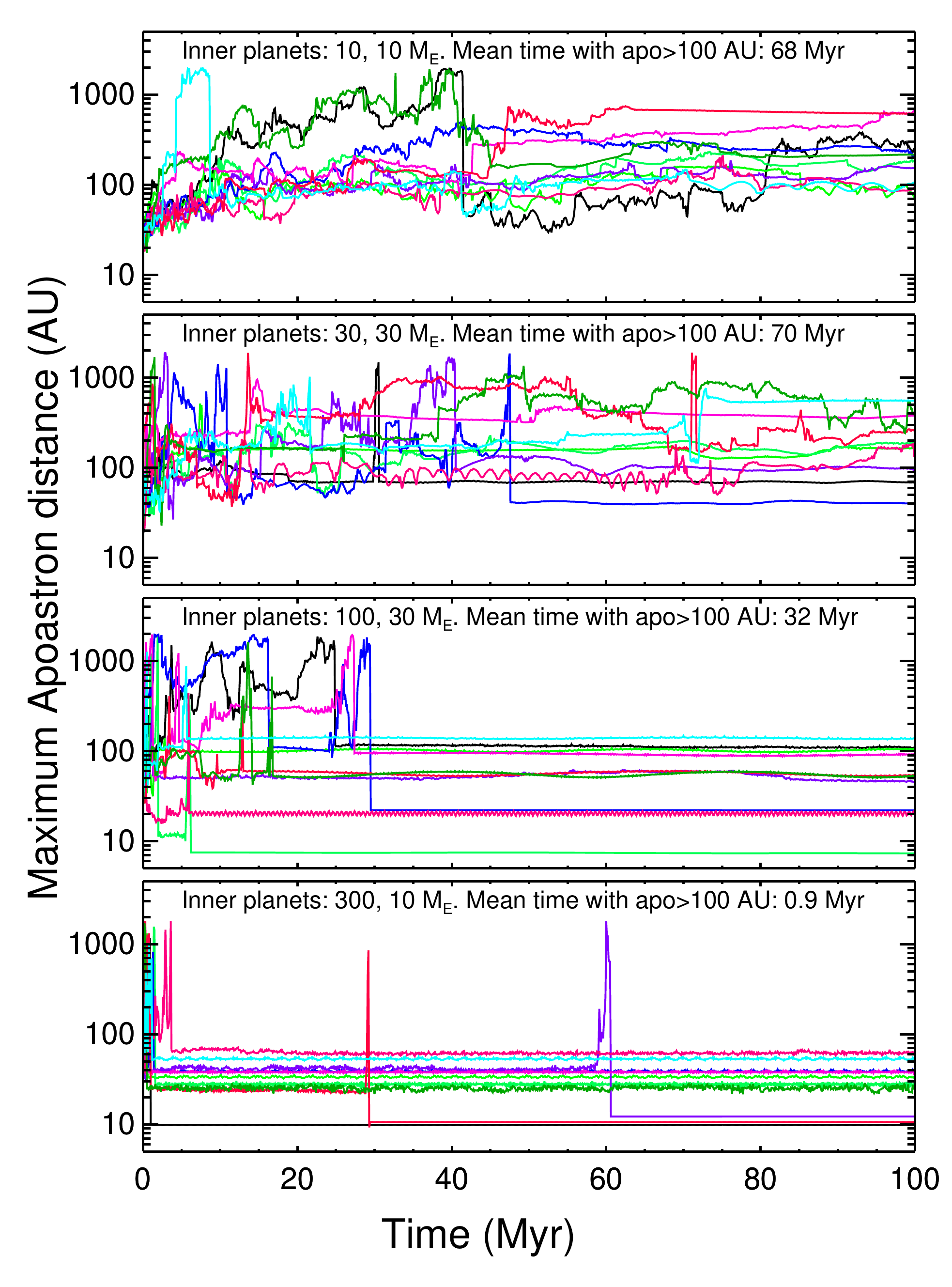}
  \caption{Evolution of the apocentre of the most distant bound planet in
    example 6-planet scattering simulations with the innermost planet at 10\,au around
    a $0.2\mathrm{\,M}_\odot$ star. Each line (10 per panel) shows the
    evolution in one simulation.}
  \label{fig:t-Q}
\end{figure}

Our flyby simulations show that capture of Novenitos can occur, 
without disrupting the cold classical KBOs, 
so long as such planets exist on orbits with apastra 
greater than 100 au around their parent star. How common are such 
low-mass, wide-orbit planets? While direct imaging surveys have revealed 
a handful of massive super-Jovian planets on very wide orbits around young stars 
\citep[e.g.,][]{Marois+10}, the occurrence 
rate of lower-mass planets on wide orbits is unknown. However, 
in regions closer to the star the planet occurrence rate is observed to 
rise strongly with decreasing planet mass 
\citep{Cumming+08,Howard+10,Fressin+13}, while microlensing 
surveys probing the snow line find that $\sim50\%$ of low-mass stars 
host a snow-line planet \citep{Gould+10,Shvartzvald+16}, 
and we might 
therefore expect suitable planets to be fairly common. 
How might 
such planets attain very wide orbits? While \emph{in situ} formation 
of super-Jovian planets may be possible via gravitational instability 
\citep{Boss97,KratterLodato16}, 
this could not lead to the formation of Neptune- or super 
Earth-mass planets. One possibility would be pebble accretion onto an 
existing core \citep{LambrechtsJohansen12}, although at distances of 10s 
or 100s of au this may require massive discs or high dust:gas ratios 
\citep{LambrechtsJohansen14}. Coagulation of small  
rocks may be possible at several hundreds of au \citep{KenyonBromley15}, 
although this process takes several Gyr, far longer than the expected 
time for which the Sun resided in its birth cluster.

More promising may be the the ejection of planets from regions closer to 
the star. Previous studies have shown that planets in the process of 
being ejected from unstable multiple systems may persist on wide 
orbits for extended periods of time \citep[][G\"otberg et al., A\&A, submitted]
{ScharfMenou09,Veras+09,Raymond+10,Malmberg+11}. 
We run scattering simulations with the hybrid integrator of the 
\textsc{Mercury} package to quantify more carefully the timescales on 
which such planets are retained on orbits that we showed above are 
suitable for capture by the Sun in a flyby. We take a $0.2\mathrm{\,M}_\odot$ primary 
and study 4- and 6-planet systems of a range of masses: The inner two planets' 
masses range from $10\mathrm{\,M}_\oplus$ to $300\mathrm{\,M}_\oplus$, 
while the outer planets are 
always assigned $10\mathrm{\,M}_\oplus$ (the mass identified by 
\citealt{BatyginBrown16}). 
Planets 
are initially started in unstable configurations 
on near-circular, near-coplanar orbits ($e<0.02$, $i<1^\circ$) separated by $3.5-5$ mutual 
Hill radii, and we conduct two sets of simulations: one ``pessimistic'' with the innermost 
planet at 3\,au and four planets in total 
and one ``optimistic'' with the innermost planet at 10\,au and 6 planets in total. 
For each set of planet masses and inner orbit, we run 10 simulations.
The systems are integrated 
for 100\,Myr. Planets are considered ``ejected'' once their distance from the 
star exceeds 10\,000\,au (In a cluster environment, the tidal field of the 
cluster or perturbations from passing stars would make themselves felt at 
these distances, \citealt{Tremaine93}.).
For each system, we record the fraction of time for which a 
$10\mathrm{\,M}_\oplus$ planet exists with an apocentre $Q$ beyond 100\,au. 
Energy is always conserved to better than 2 parts in $10^{-4}$.

Sample orbital evolution is shown in Figure~\ref{fig:t-Q}, for the 
simulations with the innermost planet at 10\,au. Systems with very massive 
planets (Saturn--Jupiter mass) swiftly eject the lower-mass planets. 
In contrast, systems comprising only $\sim$ Neptune-mass planets (10 
and $30\mathrm{\,M}_\oplus$) can retain their unstable planets for many 
10s of Myr. The mean durations for which systems retain planets with 
apocentres beyond 100\,au are indicated in Figure~\ref{fig:t-Q}, and 
are tabulated in Table~A2 of the online version of 
this paper. 
When starting 
with planets on wider orbits (innermost at 10\,au), we find that planets 
can be retained on $Q>100$\,au orbits for most of the first 100\,Myr 
(our best case being 75\%), 
while with the planets starting at 3\,au the planets can be retained for 
at most a few 10s of Myr. 
This can be attributed partly to the longer dynamical 
time-scales for the wider systems, and partly to the larger number of 
orbits required for ejection with lower-mass planets \cite[see e.g.][]{Raymond+10}. 
The wide-orbit scattered planets typically have pericentres of $\gtrsim10$\,au.

\vspace{-5mm}

\section{Discussion}

\label{sec:discuss}

How likely is the Sun to have picked up Planet~9 in a flyby, 
following scattering of the planet to a wide orbit around its 
original host? 
We can estimate the probability of a successful capture as 
$P_\mathrm{Planet9}=P_\mathrm{flyby} P_\mathrm{multi} P_\mathrm{unstable} 
f_\mathrm{wide} P_\mathrm{capture}$, where $P_\mathrm{flyby}$ 
is the probability of the Sun experiencing a suitable flyby, 
$P_\mathrm{multi}$ is the probability of having a multiple 
planetary system, 
$P_\mathrm{unstable}$ is the probability of said system being unstable, 
$f_\mathrm{wide}$ is the fraction 
of the cluster lifetime that such an unstable system retains a wide-orbit planet, 
and $P_\mathrm{capture}$ is the probability of capturing a 
wide-orbit planet into a suitable orbit around the Sun. We show in 
Section~\ref{sec:flyby} that $P_\mathrm{capture}\lesssim$4\%, and 
in Section~\ref{sec:scatter} $f_\mathrm{wide}\lesssim75\%$. 
Previous studies of cluster dynamics show that 
$P_\mathrm{flyby}\sim1$ \citep{Malmberg+07,Malmberg+11}. The 
most difficult numbers to estimate are $P_\mathrm{unstable}$ 
and $P_\mathrm{multi}$.
An optimistic estimate 
draws a parallel with Jovian planets where a very high 
incidence of instability is required to explain the eccentricity 
distribution: \cite{JuricTremaine08} find $P_\mathrm{unstable}=75\%$, 
while \cite{Raymond+11} find 83\%. We may then combine this with 
the microlensing estimate of $50\%$ of stars having a wide-orbit Neptune 
\citep{Shvartzvald+16}, and assume that all such systems are or were 
multiple. This gives an optimistic $P_\mathrm{Planet9}=$1\%.  
Pessimistically, we may assume that all multiple-Neptune systems are 
intrinsically stable 
\citep[as appears to be the case for \textit{Kepler} systems,][]{Johansen+12}; 
in a cluster environment however, 
otherwise stable systems can be destabilised by encounters with other 
stars, and \cite{Malmberg+11} find that $P_\mathrm{unstable}\sim10\%$ of Solar System 
clones (otherwise stable) in a cluster will eject a planet within 100\,Myr 
as a result of a close encounter. We then take $P_\mathrm{multi}=16\%$ 
from \cite{Gould+10}, which was based on a single detection of a two-planet 
system. Taking a pessimistic $P_\mathrm{capture}=1\%$, we then find a 
pessimistic $P_\mathrm{Planet9}\sim0.01\%$. 
Thus, the  probability for 
our Planet~9 capture scenario is $P_\mathrm{Planet9}\sim0.01-1\%$, although if 
the additional constraint on Planet~9's inclination is demanded \citep{BrownBatygin16}, 
these probabilities would shrink by a factor of 10. These
numbers compare favourably to the probability of a random 
alignment of $7\times10^{-5}$ estimated by \cite{BatyginBrown16}. 
We caution that we are calculating the conditional 
probability that Planet~9 ends up on a suitable orbit given the 
capture hypothesis, and as Planet~9's possible orbit gets refined this 
probability will become arbitrarily small. Calculation of the 
more interesting (Bayesian) probability that the capture occurred, given Planet~9's 
orbit, will have to await further studies that calculate the 
probability of Planet~9's orbit given different formation scenarios.
We also note that the study of \cite{LiAdams16} found a probability 
of $\lesssim2\%$ for the capture of Planet~9, assuming the existence 
of such a body on a wide circular orbit around another star, and 
sampling the mass of the original host from the stellar IMF. This 
number corresponds to our $P_\mathrm{flyby}P_\mathrm{capture}$; thus 
our studies are broadly in agreement. A better knowledge of the 
occurrence rate of low-mass wide-orbit planets will be needed to refine 
the probabilities for capture. 

\begin{figure}
  \includegraphics[width=.5\textwidth]{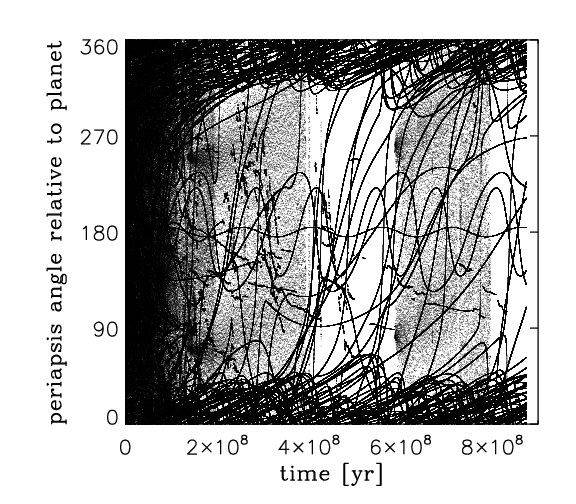}
  \caption{Orbital evolution of outer Solar System bodies scultped 
    by a captured Planet~9. We plot every 100 years 
    the longitude of periapsis of scattered disc particles between 
    70 and 1000\,au, relative to that of the planet at 283\,au. 
    After removal of unstable particles after $\sim100$\,Myr, a 
    concentration of particles at $\Delta\varpi\approx0^\circ$ is evident,
    together with a smaller number in the anti-aligned configuration
    ($\Delta\varpi\approx180^\circ$) postulated by 
    \protect\cite{BatyginBrown16}
    for the orientation of Sedna and similar TNOs. Clouds of points 
      represent highly chaotic trajectories.}
  \label{fig:longterm}
\end{figure}

How might the capture scenario be confirmed or refuted? Different histories 
for Planet 9 may affect the distributions of orbital elements of 
distant TNOs in different ways. As an example, we ran a capture simulation 
in which a $10\mathrm{\,M}_\oplus$ planet was captured from a 
$0.2\mathrm{\,M}_\odot$ star onto an orbit of 283\,au. In this simulation, 
the Sun already possesses a scattered disc of 750 test particles with 
pericentres of 40\,au and eccentricities of 0 to 0.9, and all bodies are coplanar. Following the flyby, 
the system was integrated for over 800\,Myr. The time evolution of the 
longitudes of periapsis of particles, relative to that of the planet, is 
shown in Figure~\ref{fig:longterm}. Only particles with semi-major axes 
between 70 and 1000 au 
are shown. \cite{BatyginBrown16} show that families of aligned and 
anti-aligned particles can exist under the influence of Planet~9, and in our 
integration a strong concentration of particles in orbits aligned with the 
planet is evident, together with a small number in an anti-aligned 
configuration. Furthermore, \cite{Jilkova+16} show that if multiple bodies 
are captured in an encounter then they typically have similar arguments 
of periapsis, so if the Sun picked up planetesimals along with Planet~9 
\citep[perhaps from the ``mini Oort clouds'' that can accompany 
  planet--planet scattering,][]{RaymondArmitage13}, these would add to the 
aligned population. If Planet~9 should truly exist, the capture scenario 
would thus seem to predict a much larger population of bodies with 
periapsides opposite those of Sedna and its ilk. While we caution that 
this is based on one single example of a coplanar capture, and we have neglected 
the effects of the known planets, it is likely that the distribution of 
orbital elements of distant TNOs will in the future prove a powerful 
discriminant between different scenarios for emplacing Planet~9 on its 
orbit, such as capture, \emph{in situ} formation, and scattering by the 
Solar System's known giant planets, and we encourage further dynamical 
studies to explore these possibilities.

\vspace{-5mm}

\section*{Acknowledgements}

AJM and MBD acknowledge funding from the Knut and Alice 
Wallenberg Foundation. 
SNR thanks the Agence Nationale pour la Recherche for 
support via grant ANR-13-BS05-0003-002 (MOJO).
We are indebted to Daniel Carrera for a social 
media post that led to the formation of this collaboration.
The authors wish to thank the anonymous referee for a 
swift report that improved the paper.

\vspace{-5mm}

\bibliographystyle{mnras}
\bibliography{planet9}

\appendix

\onecolumn

\section{Online tables}

\begin{longtable}{cccccccc}
    \caption{Numbers of particles (out of 750) captured in our flyby
    simulations. First three columns give the flyby parameters.
    Fourth column gives the particles' initial semi-major axes. Fifth and
    sixth columns show $n_\mathrm{capt}$, the number of particles captured
    by the Sun, and $n_\mathrm{Novenito}$, the number captured onto orbits
    suitable for Planet~9, for our simulations where
    particles had initial pericentres of 1\,au. Seventh and eighth columns
    show the same for initial particle pericentres of 10\,au. A ``-''
    indicates that the simulation was not run with these parameters.}
    \label{tab:captures}\\
      \multicolumn{4}{c}{} & \multicolumn{2}{c}{$q=1$\,au} & \multicolumn{2}{c}{$q=10$\,au}\\
      $M_\star \,[\mathrm{M}_\odot]$ & $b \,[\mathrm{au}]$ & $v_\mathrm{inf} \,[\mathrm{au\,d}^{-1}]$
      & $a \,[\mathrm{au}]$ & $n_\mathrm{capt}$ & $n_\mathrm{Novenito}$ & $n_\mathrm{capt}$
      & $n_\mathrm{Novenito}$\\
      \hline
      \endfirsthead

      \multicolumn{8}{c}{\tablename \thetable --- continued}\\
      \multicolumn{4}{c}{} & \multicolumn{2}{c}{$q=1$\,au} & \multicolumn{2}{c}{$q=10$\,au}\\
      $M_\star \,[\mathrm{M}_\odot]$ & $b \,[\mathrm{au}]$ & $v_\mathrm{inf} \,[\mathrm{au\,d}^{-1}]$
      & $a \,[\mathrm{au}]$ & $n_\mathrm{capt}$ & $n_\mathrm{Novenito}$ & $n_\mathrm{capt}$
      & $n_\mathrm{Novenito}$\\
      \hline
      \endhead

      0.1 &  500 & $1\times10^{-3}$ &  50 &  46 &  0 &  44 &  0 \\
      0.1 &  500 & $1\times10^{-3}$ & 100 &  51 &  3 &  58 &  3 \\
      0.1 &  500 & $1\times10^{-3}$ & 200 &  32 &  5 &  26 &  2 \\
      0.1 &  500 & $1\times10^{-3}$ & 400 &  11 &  0 &  11 &  3 \\
      0.1 &  500 & $1\times10^{-3}$ & 800 &   4 &  1 &   3 &  2 \\

      0.1 &  750 & $1\times10^{-3}$ &  50 &   0 &  0 &   0 &  0 \\
      0.1 &  750 & $1\times10^{-3}$ & 100 &   1 &  0 &   6 &  0 \\
      0.1 &  750 & $1\times10^{-3}$ & 200 &  81 &  7 &  83 &  3 \\
      0.1 &  750 & $1\times10^{-3}$ & 400 &  30 &  5 &  37 &  2 \\
      0.1 &  750 & $1\times10^{-3}$ & 800 &  20 &  1 &  25 &  3 \\

      0.1 &  750 & $5\times10^{-4}$ &  50 & 201 &  0 & 182 &  0 \\
      0.1 &  750 & $5\times10^{-4}$ & 100 & 178 &  3 & 150 &  5 \\
      0.1 &  750 & $5\times10^{-4}$ & 200 &  68 &  2 &  96 &  8 \\
      0.1 &  750 & $5\times10^{-4}$ & 400 &  51 &  3 &  40 &  2 \\
      0.1 &  750 & $5\times10^{-4}$ & 800 &  14 &  0 &  14 &  0 \\

      0.1 & 1000 & $1\times10^{-3}$ & 400 &   - &  - &   2 &  0 \\
      0.1 & 1000 & $1\times10^{-3}$ & 800 &   - &  - &   1 &  0 \\

      0.1 & 1000 & $5\times10^{-4}$ &  50 &  20 &  0 &  25 &  0 \\
      0.1 & 1000 & $5\times10^{-4}$ & 100 & 205 & 17 & 192 & 20 \\
      0.1 & 1000 & $5\times10^{-4}$ & 200 &  96 & 21 & 131 & 25 \\
      0.1 & 1000 & $5\times10^{-4}$ & 400 &  50 &  5 &  50 & 11 \\
      0.1 & 1000 & $5\times10^{-4}$ & 800 &  19 &  1 &  17 &  1 \\

      0.1 & 1250 & $1\times10^{-3}$ & 400 &   - &  - &   1 &  0 \\
      0.1 & 1250 & $1\times10^{-3}$ & 800 &   - &  - &   3 &  0 \\

      0.1 & 1250 & $5\times10^{-4}$ &  50 &   0 &  0 &   0 &  0 \\
      0.1 & 1250 & $5\times10^{-4}$ & 100 & 183 & 21 & 195 & 28 \\
      0.1 & 1250 & $5\times10^{-4}$ & 200 & 111 & 23 & 121 & 17 \\
      0.1 & 1250 & $5\times10^{-4}$ & 400 &  65 & 10 &  46 &  6 \\
      0.1 & 1250 & $5\times10^{-4}$ & 800 &  17 &  3 &  14 &  0 \\

      0.2 &  500 & $1\times10^{-3}$ &  50 &  22 &  0 &  34 &  0 \\
      0.2 &  500 & $1\times10^{-3}$ & 100 &  72 &  0 &  54 &  2 \\
      0.2 &  500 & $1\times10^{-3}$ & 200 &  39 &  5 &  35 &  1 \\
      0.2 &  500 & $1\times10^{-3}$ & 400 &  14 &  3 &  20 &  3 \\
      0.2 &  500 & $1\times10^{-3}$ & 800 &   - &  - &   6 &  0 \\

      0.2 &  750 & $1\times10^{-3}$ &  50 &   0 &  0 &   0 &  0 \\
      0.2 &  750 & $1\times10^{-3}$ & 100 &   0 &  0 &   6 &  0 \\
      0.2 &  750 & $1\times10^{-3}$ & 200 &  18 &  1 &  16 &  1 \\
      0.2 &  750 & $1\times10^{-3}$ & 400 &  16 &  6 &  20 &  3 \\
      0.2 &  750 & $1\times10^{-3}$ & 800 &  12 &  0 &   3 &  1 \\

      0.2 &  750 & $5\times10^{-4}$ &  50 & 191 &  0 & 129 &  0 \\
      0.2 &  750 & $5\times10^{-4}$ & 100 & 128 &  1 & 135 &  4 \\
      0.2 &  750 & $5\times10^{-4}$ & 200 &  75 &  6 &  75 &  3 \\
      0.2 &  750 & $5\times10^{-4}$ & 400 &  99 &  6 &  88 &  1 \\
      0.2 &  750 & $5\times10^{-4}$ & 800 &  54 &  1 &  74 &  2 \\

      0.2 & 1000 & $1\times10^{-3}$ & 800 &   - &  - &   7 &  1 \\

      0.2 & 1000 & $5\times10^{-4}$ &  50 &   0 &  0 &   1 &  0 \\
      0.2 & 1000 & $5\times10^{-4}$ & 100 & 153 &  8 & 149 & 16 \\
      0.2 & 1000 & $5\times10^{-4}$ & 200 & 104 & 22 & 115 & 10 \\
      0.2 & 1000 & $5\times10^{-4}$ & 400 &  54 &  8 &  53 &  2 \\
      0.2 & 1000 & $5\times10^{-4}$ & 800 &  39 &  3 &  39 &  2 \\

      0.2 & 1250 & $1\times10^{-3}$ & 800 &   - &  - &   1 &  0 \\

      0.2 & 1250 & $5\times10^{-4}$ & 100 &  67 & 14 &  70 & 15 \\
      0.2 & 1250 & $5\times10^{-4}$ & 200 & 130 & 19 & 115 & 13 \\
      0.2 & 1250 & $5\times10^{-4}$ & 400 &  72 &  5 &  54 &  7 \\
      0.2 & 1250 & $5\times10^{-4}$ & 800 &  29 &  1 &  28 &  4 \\

      0.5 &  750 & $5\times10^{-4}$ &  50 & 139 &  1 & 108 &  3 \\
      0.5 &  750 & $5\times10^{-4}$ & 100 &  84 &  2 &  86 &  3 \\
      0.5 &  750 & $5\times10^{-4}$ & 200 & 105 &  1 & 101 &  8 \\
      0.5 &  750 & $5\times10^{-4}$ & 400 & 258 & 15 & 242 & 18 \\
      0.5 &  750 & $5\times10^{-4}$ & 800 & 331 &  7 & 328 &  3 \\

      0.5 & 1000 & $1\times10^{-3}$ & 800 &   - &  - &  21 &  0 \\

      0.5 & 1000 & $5\times10^{-4}$ &  50 &   0 &  0 &   0 &  0 \\
      0.5 & 1000 & $5\times10^{-4}$ & 100 & 132 &  9 & 113 &  9 \\
      0.5 & 1000 & $5\times10^{-4}$ & 200 &  80 & 15 &  91 & 17 \\
      0.5 & 1000 & $5\times10^{-4}$ & 400 & 107 & 14 &  99 & 13 \\
      0.5 & 1000 & $5\times10^{-4}$ & 800 & 157 &  5 & 164 &  5 \\

      0.5 & 1250 & $1\times10^{-3}$ & 800 &   - &  - &   4 &  1 \\

      0.5 & 1250 & $5\times10^{-4}$ & 100 &  45 &  4 &  34 &  3 \\
      0.5 & 1250 & $5\times10^{-4}$ & 200 &  94 & 14 &  79 &  8 \\
      0.5 & 1250 & $5\times10^{-4}$ & 400 &  63 & 11 &  69 & 12 \\
      0.5 & 1250 & $5\times10^{-4}$ & 800 &  73 &  5 &  86 &  5 \\

      1.0 & 1000 & $1\times10^{-3}$ & 800 &   - &  - &  28 &  2 \\

      1.0 & 1000 & $5\times10^{-4}$ &  50 &   1 &  0 &   6 &  0 \\
      1.0 & 1000 & $5\times10^{-4}$ & 100 &  79 &  2 &  52 &  1 \\
      1.0 & 1000 & $5\times10^{-4}$ & 200 &  54 & 10 &  57 &  5 \\
      1.0 & 1000 & $5\times10^{-4}$ & 400 &  83 &  3 &  92 &  8 \\
      1.0 & 1000 & $5\times10^{-4}$ & 800 & 191 &  9 & 193 &  6 \\
      1.0 & 1250 & $1\times10^{-3}$ & 800 &   - &  - &  15 &  0 \\

      1.0 & 1250 & $5\times10^{-4}$ & 100 &  64 &  5 &  43 &  3 \\
      1.0 & 1250 & $5\times10^{-4}$ & 200 &  64 &  2 &  58 & 12 \\
      1.0 & 1250 & $5\times10^{-4}$ & 400 &  60 &  9 &  54 &  7 \\
      1.0 & 1250 & $5\times10^{-4}$ & 800 & 105 &  1 &  92 &  2 \\

      1.5 & 1250 & $1\times10^{-3}$ & 800 &   - &  - &  15 &  0 \\

      1.5 & 1250 & $5\times10^{-4}$ & 100 &  33 &  3 &  40 &  3 \\
      1.5 & 1250 & $5\times10^{-4}$ & 200 &  56 &  6 &  48 & 11 \\
      1.5 & 1250 & $5\times10^{-4}$ & 400 &  26 &  3 &  23 &  2 \\
      1.5 & 1250 & $5\times10^{-4}$ & 800 &  65 &  5 &  82 &  4
\end{longtable}

\newpage

\begin{table*}
  \begin{center}
    \caption{Mean durations for which systems retain planets
      with apastron distances $>100$\,au. First column: masses of
      inner two planets. Second column: mean duration in our pessimistic
      case (4 planets total; innermost at 3\,au). Third column:
      mean duration in our optimistic case (6 planets total; innermost
      at 10\,au).}
    \label{tab:scatter}
    \begin{tabular}{lcc}
      Planet mass $[\mathrm{M}_\oplus]$ & Pessimistic duration [Myr] & Optimistic duration [Myr]\\
      \hline
      10,  10  & 3.7  & 68.0\\
      30,  10  & 12.0 & 75.1\\
      30,  30  & 21.9 & 69.8\\
      100, 10  & 6.7  & 44.2\\
      100, 30  & 1.6  & 31.9\\
      100, 100 & 1.6  & 16.8\\
      300, 10  & 0.04 & 0.86\\
      300, 30  & 0.04 & 0.83\\
      300, 100 & 0.09 & 11.6\\
      300, 300 & 2.97 & 21.6
    \end{tabular}
  \end{center}
\end{table*}

\bsp    
\label{lastpage}

\end{document}